\documentclass[aps,twocolumn,preprintnumbers,groupedaddress,superscriptaddress,floatfix,tightenlines,reprint,nofootinbib,longbibliography]{revtex4-1}
\usepackage{mathrsfs,natbib}
\usepackage{verbatim}
\usepackage{enumerate}
\usepackage{graphicx,bbm,amsbsy,amsfonts,amssymb,amsmath}
\usepackage{bm}
\usepackage{hyperref}
\usepackage{booktabs}
\usepackage{gensymb}

\def\x{\mathbf{x}}
\def\k{\mathbf{k}}
\def\p{\mathbf{p}}
\def\r{\mathbf{r}}

\def\R{\mathbf{R}}
\def\v{\mathbf{v}}

\def\j{\mathbf{j}}
\def\A{\vec{A}}
\def\E{\vec{E}}
\def\B{\vec{B}}

\def\aB{a_{\rm B}}
\def\grad{\bm{\nabla}}
\let\vec=\mathbf

\newcommand{\He}[1]{{}^{#1}\textrm{He}}
\newcommand{\Rb}[1]{{}^{#1}\textrm{Rb}}
\let\oldAA\AA
\renewcommand{\AA}{\text{\normalfont\oldAA}}

\hypersetup{
    colorlinks=true,       
    linkcolor=red,          
    citecolor=blue,        
    filecolor=magenta,      
    urlcolor=blue           
}

\newcommand{\beq}{\begin{eqnarray}}
\newcommand{\eeq}{\end{eqnarray}}
\begin{document}

\title{Vortices in spin-0 superfluids carry magnetic flux}
\author{Aleksey Cherman}
\email{aleksey.cherman.physics@gmail.com}
\affiliation{School of Physics and Astronomy, University of Minnesota, Minneapolis, MN 55455}
\author{Theodore Jacobson}
\email{jaco2585@umn.edu}
\affiliation{School of Physics and Astronomy, University of Minnesota, Minneapolis, MN 55455}
\author{Srimoyee Sen}
\email{srimoyee08@gmail.com}
\affiliation{Department of Physics and Astronomy, Iowa State University, Ames, IA 50011 }
\author{Laurence G. Yaffe}
\email{yaffe@phys.washington.edu}
\affiliation{Department of Physics, University of Washington, Seattle, WA 98195 USA}

\begin{abstract}
Vortices in spin-$0$ superfluids generically carry magnetic
fields inside their cores, so that even neutral superfluid vortices
may be thought of as magnetic flux tubes.
We give a systematic analysis of this `vortex magnetic
effect' using effective field theory, clarifying earlier literature
on the subject.
Our analysis shows that in superfluid Helium-$4$ the vortex magnetic effect
may be large enough to be experimentally detectable.
\end{abstract}

\maketitle
\flushbottom

{\bf Introduction.}  
Superfluidity is an emergent phenomena observed in numerous
many-body systems, and plays a key role in cold nuclear, atomic, and molecular
systems. Superfluidity arises due to the formation of a Bose-Einstein condensate
(BEC) of bosonic electrically-neutral (quasi)particles at low temperatures,
so that a $U(1)$ particle number symmetry is spontaneously broken.
The condensing bosons may be individual atoms, as in the atomic superfluids 
$\He{4}$~\cite{KAPITZA1938,ALLEN1938} or $\Rb{87}$~\cite{Anderson198}, or
loosely-bound neutral Cooper pairs of fermions as in superfluid
$\He{3}$~\cite{PhysRevLett.28.885} or dense neutron matter \cite{2019arXiv190609641S}.
Superfluidity has many parallels with superconductivity,
with the crucial difference that in superconductors the condensing
particles are electrically charged.

Although superfluids arise from condensation of electrically neutral particles,
in nature these particles always have electrically charged constituents.
Novel electromagnetic (EM) properties of a superfluid with scalar condensates
(i.e., with vanishing spin $S$ and orbital angular momentum $L$)
have been explored in recent years,
starting with the experimental work of Rybalko and
collaborators \cite{Rybalko1,Rybalko2} in superfluid $\He{4}$,
which triggered considerable further
theoretical and experimental work
\cite{Kosevich1,Natsik1,Kosevich2,Pashitskii1,Melnikovsky1,Natsik2,
Loktev_Tomchenko1,Loktev_Tomchenko2,Natsik3,Gutliansky1,Shevchenko_Rukin1,
Shevchenko_Rukin2,Loktev_Tomchenko3,Tomchenko1,Poluektov1,Adamenko_Nemchenko1,
Shevchenko_Konstantinov1,Adamenko_Nemchenko2,Chagovets1,Chagovets2,
Shevchenko_Konstantinov2,Tomchenko2,Chagovets3,Adamenko_Nemchenko3,
Adamenko_Nemchenko4,Shevchenko_Konstantinov3,Shevchenko_Konstantinov4,Rybalko4}.
Previous theoretical analyses have applied a wide variety of
phenomenological models to explain the EM properties of liquid $\He{4}$ and
other superfluids.  

Our goal in this paper is to provide a systematic description of the EM
properties of scalar superfluids, i.e., superfluids with scalar order parameters,
in three spatial dimensions using the technique of effective field theory, see e.g. Refs.~\cite{Georgi:1994qn,Braaten:1996rq}.
As an application, we focus on the magnetic properties of superfluid vortices.
We will show that generic scalar superfluids embody a ``vortex magnetic effect'' (VME),
namely superfluid vortices carry non-zero magnetic flux.  We 
compare our prediction for the magnitude of the VME in superfluid $\He{4}$ to
prior estimates of this effect.

The claim that superfluid vortices carry magnetic flux might seem
surprising.
After all, the parallel statement for superconductors
holds because the superconducting order parameter is electrically charged,
implying that the superflow around a vortex
necessarily produces an azimuthal electric current and generates magnetic flux.
In contrast, the defining feature of superfluids is that the superfluid order
parameter is electrically neutral, so why should a superfluid vortex carry any
magnetic flux? 
For superfluids whose order parameters have non-vanishing spin
or orbital angular momentum,
standard orbital or spin-orbit interaction terms drive the appearance of magnetic
flux \cite{PhysRevD.25.967,PhysRevLett.51.1362}.
But for superfluids associated
with scalar order parameters, it is far less obvious why vortices should carry
any magnetic flux. 

We first discuss the underlying phenomena which produce
the dominant contribution to the VME, and other EM properties,
in dilute scalar superfluids, following Kosevich \cite{Kosevich2}.
We then present a more general effective field theory (EFT)
analysis of the problem.
One key result is that the effective Lagrangian of a generic
scalar superfluid contains an operator proportional to
$ \bm{\omega} \cdot \vec{B} $,
coupling the fluid vorticity 
$\bm{\omega} \equiv \bm\nabla \times \vec u$ to the magnetic field $\vec{B}$.
This term leads to magnetic fields localized on superfluid
vortices.

The dimensionless diluteness parameter $\gamma \equiv 2\pi n a^3$ organizes
contributions in our EFT analysis. Here $n$ is the particle density and $a$
characterizes the physical size of the particles comprising the fluid. Trapped
atomic gases can reach superfluidity while remaining dilute, $\gamma \ll 1$,
whereas $\gamma$ approaches $O(1)$ in dense fluids. 
Our EFT analysis is under complete
theoretical control for dilute scalar superfluids, and even in not so dilute
systems such as liquid $\He{4}$ the EFT approach may be used to estimate the
magnitude of the VME.

{\bf Underlying physics.}
Following Kosevich \cite{Kosevich2},
the charge density of a spherically symmetric neutral atom
may be expressed as the Laplacian of a radial function 
with rapid fall-off.
Let $a$ denote the charge radius of the atom
(i.e., root mean square radius weighted by the total charge density),
so that the Fourier transformed charge density
$\widetilde \rho(\vec k) = Z e a^2 \vec k^2/6 + O(\vec k^4)$,
with $-e$ the electron charge.
Then one may write
$\rho(\vec r) = -\nabla^2 (\frac{Z e a^2}{6} f(\vec r))$
with $f(\vec r)$ a rapidly decreasing spherically symmetric function 
which integrates to unity, or in other words
a smeared-out 3D delta function.
For hydrogen,
$a = \sqrt 3 \,\aB$ and
$
    f(\vec r) =
    (2\pi)^{-1} \aB^{-3} \, (1 {+} \aB/ |\vec r|) \, e^{-2|\vec r|/\aB}$,
with $\aB$ the Bohr radius.
Consequently, every atom generates an electrostatic potential proportional
to this smeared-out delta function.
For an arbitrary collection of widely-separated identical atoms at positions
$\{ \vec x_i \}$, the net electrostatic potential is
\begin{align}
  \Phi(\vec x) = \frac {Ze \, a^2}{6\epsilon_0} \sum_i f(\vec x{-}\vec x_i)
  \simeq \frac {Ze \, a^2}{6\epsilon_0} \, n(\vec x) \,.
\end{align}
The last form, with $n(\vec x)$ the number density of atoms,
is valid whenever the potential is to be integrated against functions
slowly varying on the scale of $a$,
so that the atomic scale details of $f(\vec x)$ are irrelevant.

The electric field experienced by a test charge is $-\bm\nabla \Phi$. An
inhomogeneous density distribution which is averaged over a region of size
$\lambda \gg n^{-1/3} \gg a$ induces a polarization $\vec P = \tfrac 16 \, Z e
\, a^2 \, \bm\nabla n$. This phenomenon has been termed \emph{flexo-electricity}
in e.g.~Ref.~\cite{Natsik1}.
If the medium is moving with some velocity $\vec v$
(small compared to the speed of light $c$),
relativistic invariance of electromagnetism implies that there will
be a magnetization $\vec M = \vec P \times \vec v$.

A superfluid vortex directly embodies the above phenomena.
For a minimal circulation vortex,
the superfluid velocity field $\vec v_{\rm s} = (\hbar/M) \, \hat{\bm \theta} / r$,
with $r$ the distance from the vortex core and $M$ the
condensing particle mass.
This leads to a total magnetic flux
\begin{equation}
   \Phi_B
    \equiv
    \mu_0 \!\int \! d\bm{\Sigma} \cdot \vec M
    =
    Z\alpha \,
    \lambda_C \, a^2 \Delta n \,
    \tfrac{2}{3} \Phi_0  ,
\label{eq:flexoresult}
\end{equation}
where $\alpha \equiv e^2 / (4\pi \epsilon_0 \hbar c)$
is the fine structure constant,
$\lambda_C \equiv 2\pi \hbar / (M c)$ is the
Compton wavelength of the fluid particles,
$\Delta n \equiv \bar n - n(0)$ is the 
difference of the average particle density $\bar n$
and the reduced density $n(0)$  at the vortex core,
and $\Phi_0 = \pi \hbar/e$ is the usual magnetic flux quantum.

{\bf Effective field theory.}  
There are further mechanisms which can generate non-vanishing
magnetic flux in superfluid vortices, including
van der Waals induced polarization in the presence of
non-uniform density,
and inertial effects in accelerating (or rotating) systems.
These mechanisms have been discussed and analyzed in various ways in
Refs.~\cite{Kosevich1,Natsik1,Kosevich2,Pashitskii1,Melnikovsky1,Natsik2,
Loktev_Tomchenko1,Loktev_Tomchenko2,Natsik3,Gutliansky1,Shevchenko_Rukin1,
Shevchenko_Rukin2,Loktev_Tomchenko3,Tomchenko1,Poluektov1,Adamenko_Nemchenko1,
Shevchenko_Konstantinov1,Adamenko_Nemchenko2,Chagovets1,Chagovets2,
Shevchenko_Konstantinov2,Tomchenko2,Chagovets3,Adamenko_Nemchenko3,
Adamenko_Nemchenko4,Shevchenko_Konstantinov3,Shevchenko_Konstantinov4,Rybalko4}.
In dilute systems these other mechanisms lead to effects
suppressed by additional factors of the small parameters $n a^3$ and/or
$m_e/M$ relative to the flexo-electric mechanism described above.
But to be confident one has not neglected some subtle yet important
physical effect,
it is very helpful to treat the problem systematically,
without the need to consider individual microscopic mechanisms in isolation.
This is the \emph{raison d'\^etre} of the effective field theory approach. 

Consider a translation-invariant system of scalar ($S = L = 0$)
electrically neutral non-relativistic bosons of mass $M$
interacting via short-range interactions.
In addition to a conserved particle number,
we assume that the interactions are also
parity and time reversal invariant.
If the system is dilute, meaning that $\gamma \ll 1$, then the
effects of interactions can be systematically characterized using
effective field theory (EFT).

A complex scalar field $\phi$ serves as a boson annihilation operator,
with the $U(1)$ particle-number symmetry acting as
$\phi \to e^{i\alpha} \phi$.
The electric and magnetic fields
are related to the electromagnetic potentials
$A_0 \equiv \Phi/c$ and $\A$ in the usual manner,
$\vec E = - \bm\nabla \Phi - \partial_t \vec A $
and $\vec B = \bm\nabla \times \vec A$.
On the low energy scales of interest,
we assume that
the only relevant degrees of freedom are those described by the
complex scalar field $\phi$ together with the electromagnetic field. 
Consequently, an action built from local gauge-invariant combinations of
these fields can provide an effective description of the system.

Since our goal is to understand EM effects, it will be helpful to take
into account the constraints of Lorentz invariance within our
non-relativistic EFT.
To the order to which we will work, it is
sufficient to demand that our EFT be invariant under linearized
Lorentz transformations representing boosts by some velocity $v \ll c$.
Such transformations act on the fields as 
$\phi(\x,t) \to e^{-i M \vec v\cdot\x}\, \phi(\x',t')$, 
$\vec E(\x,t) \to \vec E(\x',t') + \vec v\times\vec B(\x',t')$ and
$\vec B(\x,t) \to \vec B(\x',t') - \vec v\times\vec E(\x',t')/c^2$,
where $\x' \equiv \x+\mathbf{v}\, t$ and
$t' \equiv t+\mathbf{v}\cdot\x/c^2$.

Although the particles (atoms) created by $\phi^\dagger$ are neutral, they
contain charged constituents and interact with EM fields through non-minimal
couplings. To describe these interactions we assume, for simplicity, that
the $s$-wave scattering length characterizing particle collisions is comparable
to the charge radius $a$ relevant for EM interactions, with both an $O(1)$
factor times the Bohr radius $a_B$.
%
%
To enable a systematic treatment we initially assume that the system is dilute,
$\gamma = 2\pi n a^3 \ll 1$, and subsequently discuss implications when
extrapolating to liquid helium with $\gamma = O(1)$.

Under these assumptions, the most general effective action will contain
a sum of all local terms,
consistent with our symmetries,
built from $\phi$, $\E$, and $\B$ and their spatial derivatives.
The result may be expressed as
\begin{align}
S = S_{\phi} + S_{\rm EM} + S_{\phi, \,\textrm{EM}}\,,
\label{eq:EFT}
\end{align}  
where $S_\phi$ and $S_{\rm EM}$ contain the free kinetic terms plus
self-interactions of the scalar and EM fields, respectively, while $S_{\phi,
\,\textrm{EM}}$ describes the couplings between these fields. The EFT
(\ref{eq:EFT}), correctly constructed, will reproduce physics on sufficiently
large spatial and time scales.  Taking units where $\hbar = \epsilon_0 = 1$, the
spatial scales described by the EFT must be large compared to the atomic size
$a$, or equivalently for spatial momenta small compared to the EFT breakdown
scale $\Lambda \sim a^{-1}$.  The time scales described by the EFT
must be large compared to the inverse of the energy scale
$\textrm{min}(E_\Lambda, E_{\rm bind})$, where $E_{\Lambda} \equiv \Lambda^2/M$
is the energy associated with momentum $\Lambda$ and $E_{\rm bind} \equiv e^2
\Lambda$ is the atomic binding scale. Physically, of course, $E_{\Lambda}$ is
smaller than $E_{\rm bind}$ by a factor of $m/M \sim 10^{-4}$, where $m$ is the electron
mass.

To compare the importance of different terms, we define the scaling dimensions
of coordinates and fields as follows:
\begin{subequations}\label{eq:scalingPhi}%
\begin{align}
    [x] &= 1/Q,\quad
    [t] = M/Q^2,\quad
    [\phi] = Q^{3/2},
\\
    [\E] &= [c\B] = M^{-1/2} \, Q^{5/2},\quad
    [e^2] = [c] = Q/M \,,
\end{align}
\end{subequations}
where $Q$ is a characteristic momentum scale. Since $S$ is dimensionless,
every term in the Lagrange density must have dimensions of $Q^5/M$. It is
helpful to write each term in the effective Lagrangian in the form $c_i \,
\Lambda_i^{\alpha_i} \, E_i^{\beta_i} \, \mathcal O_i$, where $c_i$ is an $O(1)$
dimensionless coefficient, $\mathcal O_i$ is some combination of fields and
their derivatives, $\Lambda_i$ and $E_i$ are the natural ultraviolet (UV)
momentum and energy scales associated with the particular term in question, and
the exponents $\alpha_i$ and $\beta_i$ characterize the sensitivity of the
process described by $\mathcal O_i$ to the UV spatial momentum and energy
scales.  For all of the terms that we discuss below, $\Lambda_i =  1/a \equiv \Lambda$ and
$E_i$ is either $E_{\Lambda}$ or  $E_{\rm bind}$. 
%

The part of the action only involving the neutral bosons has the form
\begin{equation}
   S_{\phi}=\int dt\, d^{3}x \left[\phi^{\dagger}\left(i \partial_t 
    + \mu + \tfrac{{\nabla}^2}{2M}\right)\phi  
    -\frac{f_4 \, a}{M} \, |\phi|^4 + \cdots  \right] .
\label{eq:SneutralA}
\end{equation}
Here $\mu$ is chemical potential for particle number,
and the coefficient $f_4$ in the quartic self-interaction term
is a dimensionless $O(1)$ low-energy parameter which is determined by
demanding that the quartic interaction correctly reproduce
two-particle $s$-wave scattering,
while the ellipsis represents additional terms involving explicit derivatives
and/or higher powers of $\phi$,
whose coefficients must contain
additional powers of $a$ (or $1/\Lambda)$ to achieve
the correct dimensions.
Such higher order terms not explicitly shown in Eq.~\eqref{eq:scalingPhi}
have negligibly small effects on the long-distance physics in the limit
$2\pi n a^3 \ll 1$, making the properties of dilute systems of bosons
systematically calculable using the EFT, see e.g.~Ref.~\cite{Braaten:1996rq}. 

The kinetic terms of the EM fields are contained in $S_{\textrm{EM}}$, which takes the form
\begin{align}
    S_{\textrm{EM}} = \tfrac 12 \int dt\, d^{3}x \>
    \left(\E^2 -  c^2 \B^2 + \cdots\right).
\label{eq:gauge_field_action}
\end{align}
The ellipsis represents self-interactions of the EM fields induced by radiative effects.



To construct interaction terms coupling $\phi$ to the EM fields,
let
$
    \vec j \equiv \tfrac{i}{2M} \, ((\bm\nabla \phi^\dagger) \phi 
    				- \phi^\dagger \bm\nabla \phi)
$
denote the conserved particle number current density and
$n\equiv \phi^{\dagger}\phi$ the particle number density.
We also define the density gradient
$    {\bm\rho} \equiv \bm\nabla n 
$ and vorticity
$
    \bm\omega \equiv \bm{\nabla}\times\vec j
$.
The operators $\bm{\rho}$ and $\bm{\omega}$ will play roles analogous
to electric and magnetic dipole moment
densities, respectively.
The fields $\vec E$ and $\bm\rho$ are parity odd 
while $\vec B$, $\bm\omega$ and $\phi$ are parity even.
Under time reversal,
$\vec B$ and $\bm\omega$ are odd,
$\vec E$ and $\bm\rho$ are even,
and $\phi \leftrightarrow \phi^\dagger$.

Now we can discuss the leading interaction terms
in $S_{\phi, \,\textrm{EM}}$.
For our purposes, it will suffice to write out all symmetry-allowed
terms with two powers of the scalar field,
up to two powers of the EM fields,
and at most two spatial derivatives.
There are three such terms\footnote
    {
    One could eliminate the $b$ term by performing the electromagnetic field redefinition,
    $
	A_0 \to A_0 + \tfrac{1}{2} b \, e a^2 \phi^{\dag}\phi
	/(1 + c_E a^3 n)
    $
    and
    $
	\A \to \A + \tfrac{1}{2} b \, e a^2 \j
	/(c^2 + c_M e^4 a^3 n)
    $, but this would change the physical meaning of the $\E$ and $\B$ fields 
    in an unhelpful manner while not, of course, affecting any observable effects.
    We prefer to use the standard electric and magnetic fields,
    and hence choose to work with the action (\ref{eq:phiAaction})
    in which the $b$ term appears explicitly.
    },
\begin{align}
\label{eq:phiAaction}
    S_{\phi,\, \textrm{EM}}
    =& 
    \int dt \> d^{3}x \>
    \Bigl[
	b \, e\, a^2 \, (\bm{\rho} \cdot \E + \bm{\omega} \cdot \B)
	\nonumber
\\
    &{}+
	\tfrac 12 {c_{E}\, a^3} \,
	\left(n\, \E^2 -2\,\j \cdot (\E\times\B)\right)
\\
    &{}
	- \tfrac 12 {c_{M}\, e^4a^3} \,
	\left(n\, \B^2-\tfrac{2}{c^2} \, \j \cdot(\E\times\B)\right)
	+ \cdots
    \Bigr],
\nonumber
\end{align}
where the ellipsis stands for terms with higher powers of fields and/or
explicit time or space derivatives.
Note that terms proportional to
$\bm{\rho} \cdot \B$,
$\bm{\omega} \cdot \E$, and
$n\, \E \cdot \B$ are ruled out by our discrete symmetries.

We have organized the terms appearing in interaction action
\eqref{eq:phiAaction} so that each
line is invariant under linearized Lorentz boosts,
up to residuals suppressed by quadratic combinations of
boost velocity over $c$ and/or
field time derivatives over $Mc^2$.
(Such residual terms may be canceled by systematically
adding yet higher order terms to the action.)
Imposing boost invariance
reduces the set of independent dimensionless parameters
(or ``low-energy constants") characterizing the EFT,
at this order, to three:
$b$, $c_{E},$ and $c_{M}$, all of which will generically be 
${O}(1)$ unless the interactions in the underlying
microscopic theory are deliberately fine-tuned.\footnote{%
    If the interactions were to preserve a discrete symmetry $\mathcal{S}$
    which flips the EM potentials, $\Phi \to -\Phi$ and $\A \to -\A$,
    while leaving the neutral scalar $\phi$ unchanged, then $b=0$.
    Such a putative symmetry is not
    charge conjugation, which would also conjugate $\phi$, and is not a symmetry of the
    non-relativistic action \eqref{eq:SneutralA}.
    But if the field $\phi$ represents some composite
    particle built from oppositely-charged but otherwise identical constituents, then
    $\mathcal{S}$ would correspond to a charged-constituent permutation and could in principle 
    be a symmetry of the long-distance EFT \eqref{eq:EFT}.
    However, in physical systems of interest 
    there is, of course, no such symmetry interchanging electrons and ions.
    Therefore the $b$ term is not symmetry forbidden, and one should
    expect the coefficient $b$ to be ${O}(1)$;
    we show this explicitly in a toy model calculation in the Supplementary Materials.
}

The factors of $e$ and $a$ shown explicitly in the above three terms of
$S_{\phi,\rm EM}$ serve to render the coefficients $b$, $c_E$, and $c_M$
dimensionless. But since $e^2/c$ and $e^2 M a$ are dimensionless combinations,
these factors are not solely determined by dimensional analysis. The given
prefactors correspond to the statement that the relevant UV energy scale for
these EM interaction terms is $E_{\rm bind}$. 
 (Equivalently, these factors are
 also determined by noting that, in a homogeneous medium at rest, these
 interaction terms should be unaffected by sending $c \to \infty$ and $M \to
 \infty$.)

The $c_{E}$ and $c_{B}$ terms in $S_{\phi,\rm EM}$
,
 which are quadratic in EM fields, 
characterize the dielectric and diamagnetic linear response of the medium, so that 
$\epsilon/\epsilon_0 = 1 + c_E \, a^3 \, \bar n $ and
$\mu_0/\mu = 1 + c_M \, (e^2/c)^2 a^3 \, \bar n$.
We will show that
the $b$ term generates the rotation-induced polarization and magnetization
effects described in the introduction.

{\bf Magnetic structure of rotating superfluids.}
The $b$ term of $S_{\phi,\rm EM}$, which is linear in $\E$ and $\B$,
vanishes in any homogeneous, non-rotating system,
but generates novel effects in inhomogeneous systems
such as, in particular, low temperature superfluids with vortices.
To show this, we first consider a non-rotating zero-temperature
system described by the EFT \eqref{eq:EFT}, with repulsive
self-interactions and a negative chemical potential driving 
boson condensation,
$f_4 >0$ and $\mu>0$.
The field $\phi$ acquires a non-vanishing expectation value
with squared magnitude 
\begin{align}
  \bar\phi^2 \equiv |\langle \phi \rangle |^2 = \frac{M\mu}{2 \,a f_4} \,,
\end{align} 
to leading order in the EFT expansion. This indicates a superfluid state with
spontaneously broken $U(1)$ particle number symmetry. In dilute superfluids, $\bar\phi^2$ is close to (but less than)
the total particle density $\bar{n} = \langle n \rangle$    in the interacting  ground state.

If one rotates a superfluid sample, the particle number current density is given by the
superflow contribution from the condensate
$
    \j_{\rm s} \equiv -\bar\phi^2 \, \grad (\arg \langle\phi\rangle) /M
$.
   Rotation induces a non-zero superflow from variation in the
condensate phase,
$
    \v_{\rm s} \equiv \j_{\rm s} / \bar\phi^2
    = -\grad (\arg \langle\phi\rangle)/M
$,
with quantized circulation around vortices arising from
non-trivial winding of the phase,
$
    \mathcal C \equiv \oint d\bm{\ell} \cdot \v_{\rm s}(\x)
    =
    2\pi \nu / M
$,
with $\nu \in \mathbb Z$.  
Equivalently, the vorticity $\bm\omega$ is non-zero,
with surface integrals of vorticity counting the number of
vortices piercing the surface,
$
    \int_{\cal S} d\bm{\Sigma} \cdot {\bm\omega}
    = (2\pi /M) \, \nu
$. 

Due to the $\bm\omega \cdot \B$ interaction term of the EFT,
non-vanishing vorticity acts as a bias which drives a shift in
the magnetic field minimizing the energy.
Indeed, we can write 
$    b ea^2 \, \bm{\omega} \cdot \B
    =  b ea^2
	\left[ (\bm{\nabla} \times \bm{\omega}) \cdot \A 
	- \bm{\nabla} \cdot (\bm{\omega} \times \A) \right]
$.  
Integrating this term over some volume $\cal V$ with boundary
$\cal S \equiv \partial \cal V$
gives
$
    \int_{\cal V} \vec J_{\cal V} \cdot \A
    +
    \int_{\cal S} \vec J_{\cal S} \cdot \A
$,
with
$    \vec{J}_{\cal V} \equiv b e a^2(\bm{\nabla} \times \bm{\omega})
$,
$
    \vec{J}_{\cal S} \equiv b e a^2 \, \bm{\omega}\times \hat {\vec n}
$,
and $\hat {\vec n}$ an outward normal to the boundary.
This shows that in a rotating sample with boundary $\cal S$,
there is a volume EM current density $\vec{J}_{\cal V}$
proportional to the curl of vorticity
plus a surface EM current density $\vec{J}_{\cal S}$
directly proportional to vorticity.

{\bf Analysis of VME.}
A straight superfluid vortex with minimal winding is a field
configuration of the form
$\langle \phi(\x)\rangle = \bar\phi f(r) e^{i\theta}$,
using cylindrical coordinates $\x = (r,\theta, z)$ centered on and
aligned with the vortex.
Configurations with non-minimal winding in simply-connected regions typically
resemble lattices built from minimal-winding vortices.
The radial function $f(r)$
is determined, to leading order in density, by
solving the classical equations of motion.
It varies smoothly from 0 to 1 as $r$ ranges from 0 to $\infty$, with
asymptotics
$
    f(r) \sim r/\zeta
$
as $r \to 0$ and
$
    f(r) \sim 1 - \xi^2/r^2
$
as $r \to \infty$.  
Here
$
    \xi \equiv (4M\mu)^{-1/2} = a (8f_4\bar n\, a^3)^{-1/2}
$
is the ``healing length'' of the condensate.
In the low-density limit the core size $\zeta$ is
proportional to $\xi$, with $g \equiv \xi/\zeta = 0.412(4)$. 

The vorticity of
a minimal vortex is given by
$
    \bm{\omega} =
    {2\hbar\bar \phi^2} \, \hat{\vec z} \, f(r) f'(r) /(Mr)
$,
and its curl gives
\beq
    \vec J_{\cal V}
    =
    - \frac{2\hbar b e a^2 \bar\phi^2}{M}
    \left(\frac{f(r)f'(r)}{r}\right)' \, \hat {\bm\theta}\,.
\label{eq:vortex_current}
\eeq
Since $\bm\omega$ falls as $O(r^{-5})$ at large $r$, the surface current $\vec j$ is negligible.
A straightforward application of Amp\`ere's law
gives the magnetic field associated to a single minimal vortex,
$    \B_{\rm s}(\x)
    =
    2\mu_0 \hbar \frac{b \,e\, a^2 \bar\phi^2}{M}
    \frac{f(r)f'(r)}{r} \, \hat{\vec z}  \,
$
The magnitude of $\vec B_{\rm s}$ approaches
$2 \mu_0 \hbar \, {b e\, a^2 \bar\phi^2}/(M \zeta^2)$
at the center of the vortex,
and falls as $O(r^{-4})$ at large distance.
The resulting magnetic flux
is
\begin{align}
    \Phi_{\rm s}^{\nu=1}
    &= 2\pi \mu_0 \hbar \, \bar\phi^2\, \frac{b\,e\, a^2}{M} \,.
\label{fluxsf}
\end{align}
At least to this order in the EFT analysis,
the magnetic flux is completely
independent of the internal structure of the superfluid vortex.
More generally,
any configuration with the topology of a superfluid vortex
will necessarily carry a non-vanishing magnetic flux.  
A configuration of $\nu$ well-separated vortices will have total
flux which is just $\nu$ time the minimal value (\ref{fluxsf}).

Comparing our EFT result (\ref{fluxsf}) and the result (\ref{eq:flexoresult})
of our earlier discussion based on the flexo-electric effect, one sees that
the two expressions
(for dilute systems where $\Delta n \approx \bar n$)
coincide when the undetermined $O(1)$ EFT coefficient $b$ has the explicit value
\begin{align}
 b = Z/6
\end{align}

{\bf Outlook.}
Numerical estimates suggest that the VME may be 
experimentally observable.
First, consider a superfluid composed of bosonic atoms with atomic number $A$, so that
$m_e/M = 5.4 \times 10^{-4} A^{-1}$.
Expressing the flux carried by a unit-winding superfluid vortex
in terms of the superconducting flux quantum $\Phi_0 = \pi \hbar/e$ and 
and the Bohr radius
$\aB = \hbar/(\alpha m_e c)$,
(with fine structure constant 
$\alpha \equiv e^2/(4\pi \epsilon_0 \hbar c) \approx 1/137$),
we have
\begin{align}
    \frac{\Phi_{\rm s}}{\Phi_0}
    &=
    8 \pi \, \alpha^2 \, b  \,
    \bar n \,a^2 \aB \,
    \frac{m_e}{M}
\nonumber\\ &=
    7.2 \times 10^{-7}
    \,\frac{b}{A}
    \left(\frac{\aB}{a}\right)
\label{flux}
\end{align}
Note that we have replaced a factor of $\bar{\phi}^2$ by $\bar{n}$. The difference between these quantities is suppressed by a positive power of $\bar{n} a^3$, and we have only worked to leading order in $\bar n a^3$, so at the level of precision of our analysis we would not be justified in making a distinction between $\bar{\phi}^2$ and $\bar{n}$.  

To maximize the magnetic flux carried by a vortex and make experimental detection of the VME easier, we now estimate the size of the VME in non-dilute superfluids.   The paradigmatic example of a non-dilute spin-$0$ superfluid is superfluid $\He{4}$.  The diluteness parameter in superfluids is the particle density in units of the $s$-wave scattering length
$a_0$, and for $\He{4}$ it is $\mathcal{O}(1)$.  This means that $\He{4}$ is strongly coupled, and using \eqref{flux} with parameters relevant for $\He{4}$ has a roughly order of magnitude theoretical uncertainty.  To make the estimate, we set $Z=2$, so that $b \approx 1/3$, and note that 
 $\bar n \approx 0.022 \,
\textrm{A}^{-3}$, while the helium charge radius $a \approx \aB$, so $\bar{n}
a^3 \approx 0.0034$,  leading to the estimate 
\begin{align}
    \frac{\Phi_{\rm s}}{\Phi_0} &\approx 1 \times 10^{-10} \,. \quad \text{(superfluid Helium)}
\label{eq:fluxHelium}
\end{align}
We emphasize \eqref{eq:fluxHelium} is subject to roughly an $\mathcal{O}(10)$ theoretical uncertainty.   Nevertheless,  our result for the magnetic flux of a superfluid $\He{4}$ vortex is two orders of
magnitude larger than the earlier prediction of Ref.~\cite{Natsik3}.   The primary reason for the discrepancy is that Ref.~\cite{Natsik3} assumed that the flexoelectric effect arises primarily from Van der Waals interactions between atoms, rather than just the non-uniform spatial distribution of the individual atoms. Our EFT analysis gives a systematic demonstration that the latter physics drives the leading-order vortex magnetic effect, at least in a limit where systematic analytic calculations are possible.

One plausible approach to experimentally measuring the magnetic properties of
$\He{4}$ superfluid vortices would involve using SQUIDs (superconducting quantum
interference devices) to detect the vortex magnetic flux. Quantum-limited SQUIDs
of radius $1 \,\mu \text{m}$ and a noise in the range of $\sim 45\times 10^{-9}
\> \Phi_0/\sqrt{Hz}$ were reported in Ref.~\cite{Schmelz_2016}.  Using a SQUID
with this performance in an experiment with a measurement time of several days
should enable one to measure directly the vortex magnetic effect from a single
superfluid $\He{4}$ vortex.

{\bf Acknowledgments.}  We are very grateful to Fiona
Burnell for discussions and collaboration at the begin-
ning of this project, thank A. Kamenev, M. Pospelov,
S. I. Shevchenko, B. Shkolovskii, M. Shifman, and B. Spi-
vak for discussions. We also thank A. Andreev, B. Svis-
tunov, G. Volovik, and an anonymous referee for drawing
our attention to an error in an earlier version of this pa-
per, and acknowledge
support from U. Minn (AC), and DOE grants GR-024204-00001 (SS) and
DE-SC\-0011637 (LY).


\bibliography{small_circle}

\widetext
\appendix
\renewcommand{\theequation}{A\arabic{equation}}
\setcounter{equation}{0}
\renewcommand\thefigure{S\arabic{figure}}
\setcounter{figure}{0} 
\def\r{\mathbf{r}}

\section{Microscopic Toy Model}
\label{sec:appendix}
We consider a microscopic toy model describing neutral atoms comprised of oppositely charged constituents, which at low energies is described by an effective field theory analogous to \eqref{eq:phiAaction}. The discussion here supports the more general power-counting arguments we used to determine the combinations of scales appearing in the EFT describing the leading electromagnetic interactions of neutral atomic superfluids. 

Suppose that a neutral atom field $\phi$ represents a bound state of a
positively charged heavy nucleus of mass $m_+$ and a single electron of mass
$m_-$, described by the fields $\phi_+$ and 
$\phi_-$ respectively.  For our purposes the statistics of the charged particles
are irrelevant, and we take them to be bosons for simplicity.   The
corresponding Lagrangian is given by 
\begin{align} \label{eq:micro} 
\mathcal L =& -\frac{1}{4}F^2  + \phi^\dagger\left(i\partial_t-2\mu  + \frac{\bm\nabla^2}{2M}\right)\phi +\sum_{\pm}\phi_\pm^\dagger\left(i\partial_t-\mu \mp eA_0 + \frac{(\bm\nabla \mp \frac{ie}{c}\bm A)^2}{2m_\pm}\right)\phi_\pm 
 -\epsilon(\phi\,\phi_+^\dagger\phi_-^\dagger + \phi^\dagger\phi_+ \phi_-)\, ,
\end{align}
where $M = m_++m_-$ and $\mu$ is the chemical potential for the single $U(1)$
global symmetry under which $\phi_+,\phi_-$ have charge 1 and $\phi$ has charge
$2$. The $\epsilon$ coupling reflects the microscopic picture of the neutral
atom as a bound state of charged constituents, see e.g.
Ref.~\cite{Kaplan:1996nv} for a more general discussion of this sort of
approach in EFT. We will use Eq.~\eqref{eq:micro} to evaluate the amplitude for scattering of bound states due to a classical external electromagnetic field, and match the result to the same amplitude as computed in the EFT in Eq.~\eqref{eq:phiAaction}. 

To begin, we observe that the $\epsilon$ coupling relates the
atomic bound state $\phi$ to its constituents $\phi_+$ and $\phi_-$, and can be
understood as a contact interaction which approximates the re-summed ladder of
Coulomb exchanges between $\phi_+$ and $\phi_-$.   This means that $\epsilon$
can be related to the microscopic parameters $a, e, m_-, m_+$.

\begin{figure}[h!]
  \centering
  \includegraphics[width=.65\linewidth]{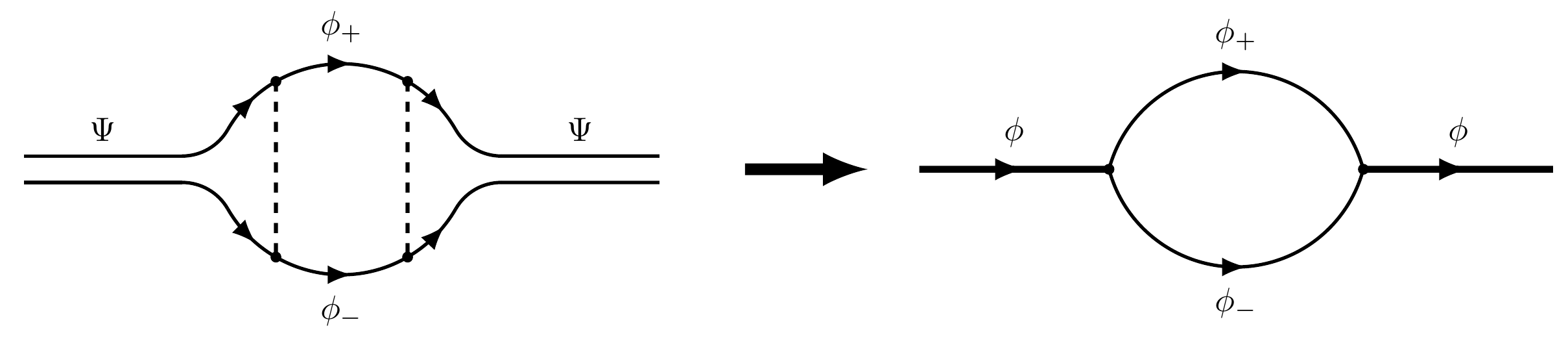}
  \caption{One-loop diagrams with and without an explicit $\phi$ field. 
  The dashed line represents Coulomb exchange, and the double line $\Psi$
   represents the result of taking all Coulomb ladder diagrams into account 
   to produce the neutral external states.  The top and bottom solid 
   lines denote $\phi_+$ and $\phi_-$ respectively. }
  \label{fig:coulomb_ladder_bound_state}
 \end{figure}

As depicted in Fig.~\ref{fig:coulomb_ladder_bound_state}, in the microscopic theory the basic bubble
self-energy diagram is approximately 
\begin{align}
\sim \int \prod_{i=1}^4 &d^4x_i \langle \Psi_{\p'}|\phi_+^\dagger(x_3)\phi_-^\dagger(x_4)\frac{e^2\delta(t_4-t_3)}{4\pi |\x_4-\x_3|} iG^-_{4,2}iG^+_{3,1}\frac{e^2\delta(t_2-t_1)}{4\pi |\x_2-\x_1|}\phi_+(x_1)\phi_-(x_2)|\Psi_{\p}\rangle\,,  \label{eq:bubble}
\end{align}
where
\begin{equation}
G^\pm_{i,j} = \int \frac{d^4k}{(2\pi)^4} \frac{e^{-i k\cdot (x_i-x_j)}}{k^0-\frac{\k^2}{2m_\pm}+\mu+i\epsilon} \label{eqn:fourier_propagator}
\end{equation}
are the non-relativistic propagators for $\phi_\pm$. To get the amplitude we are
interested in, we need to contract the external legs, taking into account that
the external states are bound states rather than free particle states.  This can
be done by writing 
\begin{equation}
\phi_+(x_1)\phi_-(x_2)|\Psi_{\p}\rangle  = e^{-ip^0 t}\phi_\p(\R_1)\, e^{-iE_b t}\psi(\r_1),
\end{equation}
where $\R_1 = \frac{m_+\x_1+m_-\x_2}{m_++m_-}$ and $\r_1 = \x_2-\x_1$ are the
center of mass and relative coordinates of the particles $\phi_{\pm}$, $p^0$ and
$\p$ are the energy and momentum of the center of mass, and $E_b$ is the binding
energy. The center of mass motion is described by a plane wave $\phi_\p(\R_1) =
e^{i\p\cdot\R_1}$. For the motion with respect to the relative coordinates, we
use the S-wave Coulomb bound state wave function of size $a$, $\psi(\r_1) \sim
\frac{1}{a^{3/2}} e^{-|\r_1|/a}$. Analogous forms hold for the final state
contraction. The contraction with the external bound states introduces a
screening length $a$ for the Coulomb interaction. Matching the momentum space
loop integral to the analogous result in the low energy effective theory amounts
to identifying
\begin{align}
\epsilon^2 \longleftrightarrow \frac{\alpha^2 c^2}{a^3}\frac{1}{(\k-\tfrac{m_r}{m_+}\p)^2+a^{-2})^2}, 
\end{align}
where $\k$ is the loop momentum and $m_r = m_+m_-/(m_++m_-)$ is the reduced
mass. The loop integral arising from Eq.~\eqref{eq:bubble} is dominated by
momentum of the order of the external momentum, so in the Coulomb propagators we
can expand in both $\p^2$ and $\k^2$ and we obtain the relation
\begin{align}
\epsilon^2 \approx \alpha^2 c^2 a \approx \frac{1}{m_-^2a}. 
\label{eq:epsilon}
\end{align}
where to get to the final expression of the right we took $m_{-} \ll m_{+}$.

\begin{figure}[h!]
  \centering
  \includegraphics[width=0.65\linewidth]{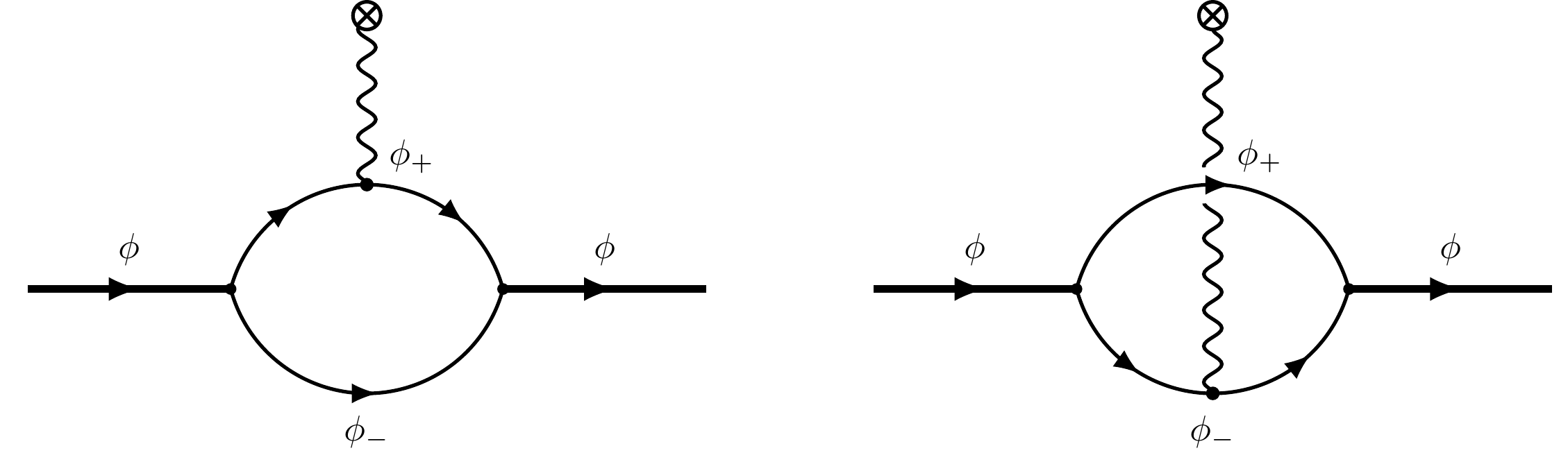}
  \caption{One-loop diagrams giving rise to the couplings between $\vec E, \vec B$ and $\bm{\rho}, \bm{\omega}$ defined in Eq.~\eqref{eq:compositeOps}. 
  The bold line denotes $\phi$ and the top and bottom solid lines 
  denote $\phi_+$ and $\phi_-$ respectively. }
  \label{fig:external_field_diagrams}
\end{figure}

We now evaluate the amplitude for scattering of neutral bound states in a
classical EM field, see Fig.~\ref{fig:external_field_diagrams}. More specifically, we expand the
S-matrix element in the external momenta and concentrate on the leading non-zero
contributions, corresponding to the leading terms in the effective Lagrangian
\eqref{eq:phiAaction}. It suffices to consider scattering from a static electric
field,  
\begin{align}
S_{fi} = &(-i)^3  e\epsilon^2 \int d^4x_1 d^4x_2 d^4x_3 \langle i | \left[\phi^\dagger(x_2)iG^+_{2,3}iG^+_{3,1}iG^-_{2,1}A_0(x_3)\phi(x_1)\right] | f \rangle -(+\leftrightarrow -)\,. 
\end{align}

The initial and final states are single-particle $\phi$ states with momenta $p$
and $p'$. Expanding in the external momenta, the leading contribution to the
amplitude is 
\begin{align}
\mathcal A \sim  e\epsilon^2 \frac{\Delta m}{m_+m_-}\left(\frac{m_r}{p^0+2\mu}\right)^{3/2}(\p-\p')^2A_0(\p'-\p)\,, 
\end{align}
where $\Delta m=m_+-m_-$. In the case of interest $m_- \ll m_+$, and to leading
order in the momentum expansion we can let $p^0 = E_b = -\frac{1}{2m_- a^2}$
equal the binding energy. Furthermore, since the chemical potential $\mu$ is suppressed relative to
$E_b$ by a factor of $m_-/m_+$, we obtain the final expression
\begin{align}
\mathcal A \sim e a^2\, (\p'-\p)^2A_0(\p'-\p),   \label{eq:Eamplitude}
\end{align}
neglecting $O(m_-/m_+)$ corrections and using Eq.~\eqref{eq:epsilon}. This
coincides with the same scattering amplitude as computed using the effective
field theory in Eq.~\eqref{eq:phiAaction} with an $O(1)$ value for the
coefficient $b$.

\end{document}